\documentclass[sigconf]{acmart}

\renewcommand\footnotetextcopyrightpermission[1]{}
\setcopyright{none}
\usepackage[english]{babel}
\usepackage[utf8]{inputenc}
\usepackage{url}
\usepackage{xspace}
\usepackage{tabularx}
\usepackage{multirow}
\usepackage{array}
\usepackage{amssymb}
\usepackage{listings}
\usepackage{enumitem}
\usepackage{subcaption}
\usepackage{listings}

\newcolumntype{Y}{>{\centering\arraybackslash}X}
\newcolumntype{C}{>{\raggedright\arraybackslash}X}

\newcommand{\rotate}[1]{\parbox[t]{2mm}{\rotatebox[origin=l]{90}{\parbox[t]{1.5cm}{#1}}}}

\newcommand*{\eg}{\textsl{e.g.,}\@\xspace}
\newcommand*{\ie}{\textsl{i.e.,}\@\xspace}
\newcommand*{\etal}{\textsl{et~al.}\@\xspace}

\newcommand{\todo}[1]{}

\newcommand{\sysname}{FlexState\xspace}

\def\sectionautorefname~#1\null{\S#1\null}
\def\subsectionautorefname~#1\null{\S#1\null}
\def\subsubsectionautorefname~#1\null{\S#1\null}

\usepackage{adjustbox}
\usepackage{array}
\newcolumntype{R}[2]{
    >{\adjustbox{angle=#1,lap=\width-(#2)}\bgroup}
    l
    <{\egroup}
}
\newcommand*\rot[2]{\multicolumn{1}{R{#1}{#2}}}

\begin{document}
\title{FlexState: Enabling Innovation in\\Network Function State Management}

\author{Matteo Pozza}
\affiliation{
  \institution{University of Helsinki}
}

\author{Ashwin Rao}
\affiliation{
  \institution{University of Helsinki}
}

\author{Diego Lugones}
\affiliation{
  \institution{Nokia Bell Labs}
}

\author{Sasu Tarkoma}
\affiliation{
  \institution{University of Helsinki}
}

\begin{abstract}
Network function (NF) developers need to provide highly available solutions with diverse packet processing features at line rate. A significant challenge in developing such functions is to build flexible software that can be adapted to different operating environments, vendors, and operator use-cases. Today, refactoring NF software for specific scenarios can take months. Furthermore, network operators are increasingly adopting fast-paced development practices for continuous software delivery to gain market advantage, which imposes even shorter development cycles. 
A key aspect in NF design is \textit{state management}, which can be optimized across deployments by carefully selecting the underlying data store. However, migrating to a data store that suits a different use-case is time consuming because it requires code refactoring while revisiting its application programming interfaces, APIs.

In this paper we introduce \sysname, a state management system that decouples the NF packet processing logic from the data store that maintains its state. The objective is to reduce code refactoring significantly by incorporating an abstraction layer that exposes various data stores as configuration alternatives. 
Experiments show that \sysname achieves significant flexibility in optimizing the NF state management across several scenarios with negligible overhead. 
\end{abstract}

\maketitle

\todo{I think there are two different problems that need to be made explicit.
a) state management systems have different APIs, so adopting a new state management system requires significant code refactoring in the network functions.
b) state management systems do not offer a mechanism to change/substitute/upgrade the internal datastore, and this is problematic because databases with new features and fixing bugs are released all the times.

FOCUS IS ON CHANGING DATASTORE, NOT ON CHANGING STATE MANAGEMENT SYSTEM.}

\section{Introduction}
\label{sec:intro}
Network functions (NFs), such as network address translators (NATs), load balancers or intrusion detection systems (IDS) are stateful entities, meaning that there exists an inherent trade-off in maintaining a consistent shared state across packet flows, or among multiple NF instances, while processing packets at line rate~\cite{Rajagopalan13SplitMerge, GemberJacobson14OpenNF}. This trade-off is even more challenging with the adoption of virtualization technologies to dynamically scale NF instances according to traffic variations.
As a consequence, state management systems are designed and optimized for a specific set of requirements, which determines the appropriate data store applicable to a specific use-case. 
For example, StatelessNF~\cite{Kablan17StatelessNF} relies on a remote key-value store (KVS) to provide reliability, while S6~\cite{Woo18S6} uses a distributed hash table (DHT) to optimize for high performance instead.

More concretely, NF operational requirements can vary quite significantly among use-cases. For instance, a network tailored for stock trading~\cite{Sun17NFP} targets the lowest achievable latency; whereas for voice and video services~\cite{Han15NFV}, networks must be robust to disruptions. Thus, the specific use-case influences the selection of the data store used internally by the state management system. In practice, developers need to design  packet-processing functions for a variety of scenarios with different data store optimized features. 

A challenge in NF development today, is that the packet processing logic is tightly coupled to the state management system. The reason for this is performance. However, changes in the NF requirements, new use-cases, or upgrading the existing data store with new features, require a coding effort that significantly delays the NF deployment in production. That is, the process of identifying all the state variables in thousands of lines of NF code and adapt them to a new data store API is error-prone and time-consuming~\cite{GemberJacobson14OpenNF, Khalid16StateAlyzr}. 

\begin{figure}
    \centering
    \includegraphics[width=\columnwidth]{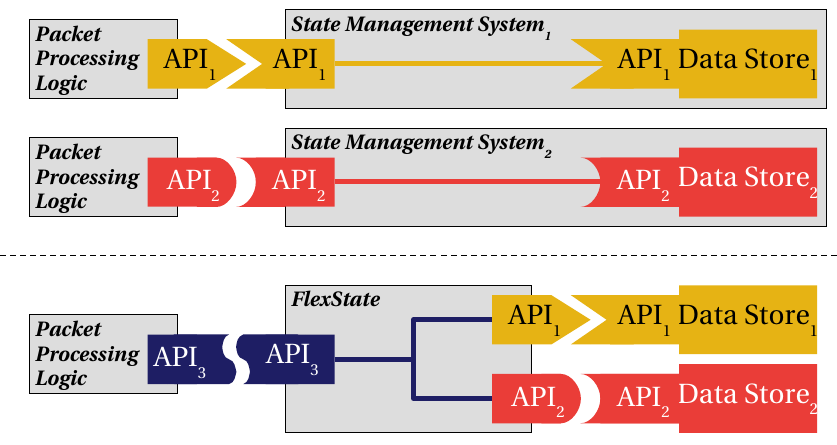}
    \caption{{\bf \sysname exemplified}.
    {\sl The APIs exposed by state management systems are tightly coupled with the data store used internally.
    \sysname exposes a single API that simplifies the adoption of a data store of choice by abstracting its implementation.}}
    \label{fig:intro}
\end{figure}

In this paper, we argue that is essential to decouple the state management from the packet processing logic to reduce development times without affecting performance. That is, reusing the code base across data stores must not require continuous refactoring but can be provided through configuration.
\autoref{fig:intro} (top) shows the dependencies between the function execution and the API exposed by the state management system. The figure also illustrates (bottom) the API translation made by \sysname to abstract the characteristics of the underlying state management systems. Conceptually, this approach has been applied in other contexts; for instance, Apache Libcloud \cite{libcloud} provides a library for interacting with many cloud service providers through a unified API. Similarly, the Serverless framework \cite{serverless} offers an open source CLI to deploy serverless apps across many platform providers.

\autoref{tab:comparison}, shows how different optimization goals influence the interface design and the data store selection. For instance, the CHC~\cite{Khalid19CHC} state management system uses a custom key-value store to support collections, such as lists, and \textit{method call shipping}, i.e., the NF can offload some operations on the collections to the data store itself. Although this feature is helpful to ensure consistency, it is not offered by other solutions like OpenNF or StatelessNF which offer complementary capabilities, meaning that developers need to engage in non-trivial code refactoring to avoid vendor lock-in.

In this paper, we introduce \sysname\footnote{Note that this work does not raise any ethical issues.} to allow state management systems to use data stores interchangeably. \sysname enables NF developers to access and manipulate the state of NFs through a single API exposed to the packet processing logic while leveraging a range of data store drivers to translate API operations into data-store-specific query language. We demonstrate in \autoref{sec:evaluation} the potential of \sysname by using two structurally different data stores without requiring any modification on the NF's packet processing code, and without incurring overhead.

The rest of the paper is organized as follows. Section  \ref{sec:background} provides background information and motivations behind \sysname, while \autoref{sec:architecture} elaborates on its design and architecture. Implementation details are discussed in \autoref{sec:implementation}. Evaluation results are presented and discussed in \autoref{sec:evaluation} and \autoref{sec:discussion}, respectively. Related work is described in \autoref{sec:related}. Finally we conclude in \autoref{sec:conclusions}.

\section{Background and Motivation}
\label{sec:background}

\begin{table}
\begin{tabular}{r | l | l l l l l l l l}
&  & \multicolumn{8}{c}{\bf API Features} \tabularnewline
\cline{3-10}
{\bf System} & {\bf \rotate{Goal}} &  \rot{25}{1em}{Get/Set} & \rot{25}{1em}{Collections} & \rot{25}{1em}{Consistency Tuning} & \rot{25}{1em}{Locking} & \rot{25}{1em}{Timers} &  \rot{25}{1em}{Merging} & &  \tabularnewline
\hline
Split/Merge \cite{Rajagopalan13SplitMerge} & S & \checkmark & $\circ$     & \checkmark & \checkmark & \checkmark & \checkmark \tabularnewline
OpenNF~\cite{GemberJacobson14OpenNF}       & A & \checkmark & $\circ$     & \checkmark & $\circ$    & $\circ$    & \checkmark \tabularnewline
StatelessNF~\cite{Kablan17StatelessNF}     & R & \checkmark & $\circ$     & $\circ$    & $\circ$    & \checkmark & $\circ$ \tabularnewline
S6~\cite{Woo18S6}                          & P & \checkmark & \checkmark  & \checkmark & $\circ$    & $\circ$    & \checkmark \tabularnewline
libVNF~\cite{Naik18libVNF}                 & S & \checkmark & $\circ$     & $\circ$    & $\circ$    & $\circ$    & $\circ$ \tabularnewline 
CHC~\cite{Khalid19CHC}                     & P & \checkmark & \checkmark  & $\circ$    & $\circ$    & $\circ$    & $\circ$ \tabularnewline 
\hline
\end{tabular}
\caption{{\bf Comparing the APIs of state management systems}. 
{\sl Each system is designed around a different goal: Accuracy~(A), Performance~(P), Reliability~(R), or Scalability~(S).
The goal drives the choice of the data store, which affects the API of the state management system; \checkmark indicates a feature is supported, while~$\circ$~indicates that a features is not supported.
\sysname is designed to address the heterogeneity of the APIs exposed by these state management systems.
}}
\label{tab:comparison}
\end{table}

\todo{Terminology of S6, CHC: NFV, NF, NF instance}

Network functions, e.g., NATs and load balancers, can be deployed in different scenarios, such as data center interconnects or enterprise networks, and for a variety of use-cases ranging from latency-sensitive to bandwidth-intensive. When deployed in production environments, NFs are expected to scale with the traffic load. For this reason there can be multiple NF instances acting on a packet flow, or sharing state across multiple flows concurrently. Furthermore, within each NF instance, developers can also parallelize the packet processing to fully utilize the available CPU resources and increase performance. In this section, we describe the implications of the NF design choices mentioned above on the state management system.

\subsection{The State of a Network Function}
\label{sec:nfstate}

NFs are stateful entities that require timely access and the ability to operate on the state of the variables used for processing the packets flows of the incoming traffic. Therefore, the state of a NF can be binned in two categories: {\em per-flow} state, and {\em cross-flow} state~\cite{Khalid19CHC}.
The per-flow category represents the state processing corresponding to packets of a specific flow. 
In contrast, the cross-flow state represents the state information considered when processing packets from all the flows traversing the NF. 
For instance, the per-flow state in a NAT NF contains the pair of IP addresses of a given TCP/UDP flow, while the cross-flow state includes the available IP addresses and port numbers that can be used when performing the translation.

\subsection{Motivation}
\label{sec:statemgmtsystems}

Maintaining the state information of NFs at line rate is challenging in terms of performance and consistency because each NF instance may have multiple threads processing packet flows, and there may be multiple NF instances in the network. For this reason, recent research have explored several state management alternatives to handle the entire life-cycle on behalf of the NFs by taking care of aspects such as consistency and correctness of data. \autoref{tab:comparison} summarizes some of these state management systems, each of which is optimized for a specific goal. Note that some systems are tailored for reliability, thus prioritizing that the NF state is always available, while other systems are designed around scalability only, thus focusing on the capability of supporting a varying traffic load in an elastic, eventually consistent, manner. 
Developers can select the data store that is most suited for the use-case(s) in scope, according to the goal for which the system is developed. For example, StatelessNF~\cite{Kablan17StatelessNF} uses an external Key-Value Store (KVS) to maximize reliability, while S6~\cite{Woo18S6} adopts a Distributed Hash Table (DHT) to offer high performance.

As mentioned in \autoref{sec:intro}, a common issue across all state management systems is that they are ultimately tightly coupled to a given data store, and do not provide any simple explicit mechanism for migrating to a different data store. This inflexibility limits the features and functionality that the state management system API offers to the packet processing logic for operating on the NF state. 
For example, the data store of CHC~\cite{Khalid19CHC} supports collections, such as lists and maps, and it allows users to offload operations on the collections to the data store, \eg incrementing the first element of a list. 
The API exposed by CHC thus supports collections and operation offloading, a feature which is not supported by the APIs of many other state management systems in \autoref{tab:comparison}. 
Similarly, StatelessNF~\cite{Kablan17StatelessNF} relies on a data store named RAMCloud to support timers, whereas S6~\cite{Woo18S6} internally uses a DHT without supporting timers. In this case, adding RAMCloud capabilities to NFs running S6 requires arduous code refactoring.
LibVNF~\cite{Naik18libVNF}, in turn, supports data locality through private local data stores for each NF instance, and a global shared data store for all NF instances. In contrast, S6 is designed to ensure that the access to the state is location independent.

Another limitation, caused by tightly coupled data stores, is the inability to streamline (in the production NF) new features, bug fixes, and performance enhancements that are constantly released by the data store developers. Network operators would highly benefit from incorporating constant data store upgrades in their state management system, but this is a challenging process~\cite{Khalid16StateAlyzr}.
Gember-Jacobson \etal~\cite{GemberJacobson14OpenNF} report that porting a carrier-grade NF such as Bro~\cite{Paxson99Bro} to their state management system required to change thousands of lines of code. High refactoring costs usually result in some form of lock-in for network operators.

The limitations mentioned above motivate our design choices to decouple the packet processing logic from the data store, and develop a flexible NF state management system that allows to alternate across multiple data store APIs by simply modifying configuration parameters. 

While decoupling reduces code refactoring when upgrading existing data stores, or migrating to new ones, it also adds a level of indirection that can affect the line rate performance, as shown in \autoref{fig:intro}. Thus, the challenge addressed in this paper is two-fold:

a) To enable flexibility in the choice of the data store used for NF state management, and 

b) To preserve scalability and line-rate performance.\\

Next, we provide more details on the \sysname system design, architecture and implementation.

\section{\sysname Architecture}
\label{sec:architecture}

In the following, we describe the key components of \sysname, namely the API and the data store drivers, and we explain how they can be used by NF developers and network operators (§\ref{sec:key_components}).
We then describe how \sysname manages state information (§\ref{sec:state_management}) and the optimizations we designed to enable NFs using \sysname to perform at line rate (§\ref{sec:perf_optimizations}).

\begin{figure*}
\centering
\includegraphics[width=\textwidth]{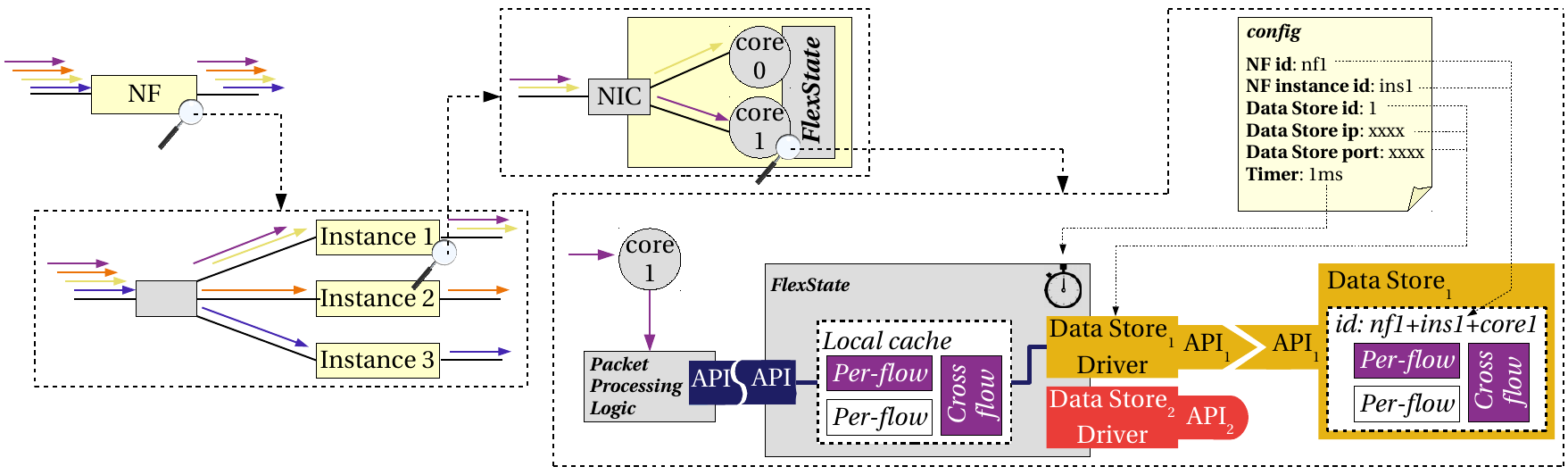}
\caption{\textbf{\sysname architecture}.
{\sl A NF comprises a variable number of instances, each of which processes a subset of flows.
Within each instance, the NIC typically applies RSS and distributes the flows across the cores made available to the NF instance.
On each core, the packet processing logic sends state operations to \sysname, which handles the state in a local cache.
\sysname uses a config file describing a) the data store driver to use and the information to reach the data store, and b) the frequency with which the state in the local cache is pushed to the data store.}}
\label{fig:partitioning}
\end{figure*}

\subsection{Key Enablers}
\label{sec:key_components}

\subsubsection{API}
\label{sec:api_network_functions}

The main goal in designing \sysname API is to provide the features required by packet processing logic of NFs.
To identify the features to be included, we use the classification of the APIs exposed by the other state management systems shown in Table~\ref{tab:comparison}.
We present here how \sysname supports the two main features, namely the support for get/set operations and collections.
We discuss how \sysname can support the remaining features in §\ref{sec:discussion}.

\sysname API provides a set of data structures, each supporting a range of operations.
Each data structure has a type, which determines the operations supported on the data structure.
Moreover, when instantiating a data structure using the \sysname API, the NF developer assigns an identifier (id) to the data structure.
Both the type and id are used to identify the data structure in the data store as explained in §\ref{sec:keying_schema}.

\paragraph{Name-value pairs and counters.}
The basic data structure provided is the name-value pair.
The NF developer can use this data structure to save a generic blob of data using a string as identifier.
The API calls exposed on such name-value pairs correspond to a Create, Read, Update, and Delete (CRUD) interface.
Note that Read and Update calls correspond to Get and Set calls in key-value stores.
In addition to name-value pairs, \sysname exposes a dedicated data structure for counters.
Indeed counters are used in a multitude of tasks, such as counting the total number of active flows, and they are natively supported by many data stores~\cite{Redis,Cassandra}.
In the \sysname API, counters expose the same CRUD calls of the name-value pairs, and they also expose the call \texttt{add(value)}, which adds the specified \texttt{value} to the current value of the counter.

\paragraph{Collections.}
Similarly to other state management systems \cite{Woo18S6,Khalid19CHC}, \sysname exposes collections, namely lists, sets, and maps.
In addition to the CRUD interface, the calls exposed by collections take inspiration from the calls exposed by the correspondent data structures in the C++ standard containers library~\cite{stdlib}.
\sysname exposes also countermaps, \ie maps whose values are counters.
They are useful in many NF tasks, such as counting the number of packets for each active flow.
Countermaps expose the same calls as a regular map, but they also expose the \texttt{addTo(key, value)} call, which adds the specified \texttt{value} to the current value of the counter identified by \texttt{key}.

\subsubsection{Data Store Drivers}
The goal of a data store driver is to translate \sysname API calls using the query language of the data store.
In this way, when changing data store, it is only needed to configure \sysname to use the appropriate data store driver because the packet processing logic of the NF is written using the \sysname API and it does not need to be changed.
An example of translating API calls is shown in Table~\ref{tab:example_queries}.
A key challenge here is to realize a simple mechanism that allows network operators to change the data store driver adopted.
In \sysname, the network operator compiles a configuration file, which is fed to the state management system.
The network operator specifies the data store driver to be adopted, as well as the parameters needed by the driver to connect to the data store, \ie IP address and port.
To specify the driver, the network operator uses a label, which identifies the data store of the driver.

When a new data store is released, a driver for the data store can be developed and added to \sysname.
Developing the driver is not hard because it is enough to implement \sysname API calls.
Once the driver is developed, the driver can be integrated in \sysname providing the label to be used in configuration file to identify the data store.
After the integration process, a network operator can setup the new data store, compile the configuration file accordingly, and \sysname will start using the new data store for state management.

\begin{table*}
    \centering
    \begin{tabularx}{\textwidth}{l|l|X}
    \textbf{Type \& Call} & \textbf{Redis} & \textbf{Cassandra} \\
    \hline
    Counter & \multirow{2}{*}{INCRBY nf1@ins1@1@Counter@counter\_id n} & \multirow{2}{=}{UPDATE nf1@ins1@1.Counter SET value = value + n WHERE key=counter\_id} \\
    add(n) & & \\
    \hline
    Map & \multirow{2}{*}{HSET nf1@ins1@1@Map@map\_id k n} & \multirow{2}{=}{INSERT INTO nf1@ins1@1.Map (key1, key2, value) VALUES (map\_id, k, n)} \\
    insert(k,n) & & \\
    \hline
    Countermap & \multirow{2}{*}{HINCRBY nf1@ins1@1@Countermap@cmap\_id k n} & \multirow{2}{=}{UPDATE nf1@ins1@1.Countermap SET value = value + n WHERE key1=cmap\_id AND key2=k} \\
    addTo(k,n) & & \\
    \end{tabularx}
    \caption{\textbf{Examples of API conversion}.
    {\sl The data store drivers use the information provided by \sysname as described in §\ref{sec:keying_schema}.
    NF id, NF instance id, and core id are taken from Figure~\ref{fig:partitioning}.
    Symbol @ is used to separate the fields.}}
    \label{tab:example_queries}
\end{table*}

\subsection{State Management}
\label{sec:state_management}

\subsubsection{Organization}
A state management system must ensure correctness of state information, for example making sure that concurrent write operations do not corrupt state.
While previous work have extensively discussed about the difficulties in handling cross-flow state~\cite{Woo18S6,Khalid19CHC}, we argue that also the handling of per-flow state is not trivial.
Indeed, the state management system must handle appropriately race conditions on state information when a NF instance runs on multiple cores.

\sysname solves this problem using \textit{partitioning}.
According to this principle, data is partitioned among a group of executors, such as NF instances or cores, and each executor accesses and modifies only its own data.
In this way, executors are made independent from each other and they do not incur race conditions because there is no shared data.
When considering multiple NF instances, \sysname divides the state information among the instances, and each instance accesses and modifies its own state information only.
As an example, when considering a cross-flow state information, such as a counter for the total number of flows traversing the NF, the counter is split into a set of independent counters, each one of them associated to a single instance.

Partitioning is not applied only across different NF instances, but also within each instance.
Figure~\ref{fig:partitioning} illustrates how \sysname applies partitioning within a single NF instance.
\sysname leverages the fact that modern Network Interface Cards (NICs) support Receiving Side Scaling (RSS).
When RSS is activated, the flows arriving at the NIC are distributed evenly among the cores made available to the NIC.
Crucially, the NIC forwards packets of the same flow always to the same core~\cite{Sherry15Rollback}.
As a consequence, for each flow, there is a single core processing its packets, so race conditions cannot occur handling per-flow state.
When considering cross-flow state instead, \sysname applies partitioning by splitting the cross-flow state among the cores allocated for the NF instance.
Considering the example of the counter for the total number of flows, \sysname splits the NF instance counter into a set of independent counters, each one of them associated to a single core.
While each counter is still cross-flow state, it is accessed and modified only by a single core, and thus race conditions cannot occur.

Note that designing partitioning-aware NFs is a non-trivial task.
NF developers need to split across cores and NF instances the state information that are typically shared.
In §\ref{sec:eval:nfs} we provide two examples of how to perform this splitting.
Moreover, network operators need additional tools to view NF state as a single entity, \eg to examine the overall load across all NF instances.
In §\ref{sec:discussion} we discuss the need of combiners~\cite{Rajagopalan13SplitMerge}, which are used to obtain a single representation of state that is scattered across NF instances.

\subsubsection{Identification}
\label{sec:keying_schema}
Each core of each NF instance manages a piece of NF state in an exclusive fashion.
Nevertheless, the NF state is stored in a data store which is shared by all cores of the NF instance.
Moreover, the data store might be shared also by other NF instances and by other NFs.
Therefore, there is a need for creating unique identifiers for state information so that partitioning can be applied in the data store.

In \sysname's configuration file, the network operator specifies two additional information, a) a NF identifier, and b) a NF instance identifier.
These information are used to distinguish data of different NFs and to distinguish data of different instances of the same NF, respectively.
\sysname also leverages the id of the core from which state operations are being issued to distinguish data used by different cores of the same NF instance.
For each data structure created using the API, \sysname creates a unique key combining together a) the NF identifier, b) the NF instance identifier, and c) the id of the core.
\sysname also combines the type of the data structure issuing the state operation, \eg counter, and the id of the data structure assigned by the NF developer.
These two pieces of information allow distinguishing data managed by the same core.
For completeness we provide an example of how a unique id is created in the Appendix.

\subsection{Performance Optimizations}
\label{sec:perf_optimizations}

\subsubsection{\textit{no\_wait} calls}
All API calls described so far return either a result of a query, \eg the data corresponding to a \textit{get} call, or an acknowledgement of completed operation.
While normally the NF waits for the reply to come back, there are situations in which waiting for the reply form the data store is not desirable.
For example, if state operations are issued in the packet processing loop, waiting for responses from the data store can slow down the NF.
Past NF state management systems solve this issue by adopting \textit{no\_wait} calls, which issue state operations without waiting a response from the data store~\cite{Woo18S6,Khalid19CHC}.
Therefore, we complement the regular API calls with no\_wait calls, which can be effectively used in the packet processing loop without slowing down the NF.
Note that not all calls are suitable for a no\_wait version.
For example, if a NF uses the \textit{get} call to obtain data from the data store, then it needs to wait the response from the data store.
In our experience, normal calls are used only in initialization or shutdown of the NF, but not within the packet processing loop.

\subsubsection{Asynchronous Updates}
\label{sec:async_updates}

The rate at which NFs process packets can be very different from the throughput of data stores, \ie number of operations per second~\cite{Chandramouli18Faster}.
In this case, the overall processing rate of the system corresponds to the rate of the slowest between the data store and the packet processing logic.
The problem stems from the idea of performing state operations on the data store every time a packet is processed in a synchronous fashion.
\sysname solves this problem by decoupling the packet processing loop from the state management operations.
The packet processing loop operates on a local cache of the state, thus avoiding the need to communicate synchronously with the data store.
A periodic operation is then issued to update the state on the data store with the changes that have been performed on the local cache.
In effect, these correspond to asynchronous updates to the data store.

A key aspect to consider is the frequency with which the update operations are issued to the data store.
Depending on the use-case, a network operator might require high availability, and thus to have very frequent updates on the data store~\cite{Rajagopalan13PicoReplication,Sherry15Rollback}. 
\sysname allows the network operator to configure the frequency of updates to the data store in the configuration file, as shown in Figure~\ref{fig:partitioning}.
More specifically, the network operator sets the time gap between updates to the data store, \eg 1~ms.
By decreasing the value, the network operator increases the frequency of updates to the data store at expenses of a higher amount of traffic between \sysname and the data store.

\section{Implementation}
\label{sec:implementation}

Our \sysname prototype consists of approximately 5K lines of C++.
In the following, we describe the tools and techniques we adopted to implement each component of \sysname architecture.

\subsection{Key Enablers}
The key goal of \sysname is to enable changing the data store driver without requiring changes in the code of the packet processing logic of the NFs nor in the state management system.
Therefore, we implemented \sysname API as an \textit{interface}, which is instantiated by the data store drivers.
\sysname internally uses the interface to issue state operations, thus remaining agnostic to the actual data store driver selected.

To exemplify how \sysname API can be used with different data stores, we implemented the drivers for a range of data stores in our \sysname prototype.
We choose Redis~\cite{Redis} and Cassandra~\cite{Cassandra} because they belong to different data store families, \ie Redis is a key-value store with a flat key space while Cassandra organizes data in tables.
Redis has been used in NF systems due to its consistency guarantees~\cite{Naik18libVNF}, while the fault tolerance capabilities of Cassandra can be leveraged with use-cases with very stringent availability requirements~\cite{Lugones17Aidops}.
Moreover, Redis and Cassandra are both carrier-grade data stores, \ie they are used and maintained by major IT companies: using carrier-grade data stores for NF state management provides further benefits, as discussed in §\ref{sec:discussion}.
We also implemented the driver for a in-memory hashmap.
The hashmap is not shared among NF instances and it executes locally to each NF instance, \ie it runs in the same host of the NF instance.
We use the hashmap only for benchmarking purposes.

Table~\ref{tab:example_queries} shows few examples of how the data structures and the API calls of \sysname are converted by the data stores drivers.
Supporting counters is straightforward because both Redis and Cassandra natively support counters, and the \texttt{add(value)} call of the \sysname API can be mapped directly to the corresponding calls in Redis and Cassandra, respectively \texttt{INCRBY} and addition operand.
Supporting maps and countermaps in Redis is easy as well because the data store supports both data structures and thus it natively exposes calls for inserting an element into a map and incrementing a value in a countermap.
With Cassandra, we implement maps and countermaps by expanding them in normal tables because the native support for maps in Cassandra is inefficient.
We discuss this aspect more in details in §\ref{sec:impl_state}.

\subsection{State Management}
\label{sec:impl_state}
\sysname leverages RSS to distribute the flows across the available cores and partitioning to avoid inter-core contention, and thus improve the performance and the scalability of the system.
In our implementation, we thus use Seastar~\cite{Seastar}, a framework that has been used successfully in other related work~\cite{Duan19NetStar}.
Seastar takes care of distributing flows across the available cores by configuring the NIC to apply RSS and linking each hardware queue of the NIC to a different core.
If the number of available cores is higher than the number of hardware queues in the NIC, then Seastar creates software queues for the remaining cores and it performs RSS in software to distribute the flows evenly among all available cores.
For each available core, Seastar creates a thread, it pins the thread to the core, and it configures the thread to process the packets of the queue linked to the core.
Seastar also facilitates partitioning by creating per-core data structures, which are accessed and modified only by the thread assigned to the core.
Lastly, Seastar natively integrates with DPDK~\cite{DPDK}, which we adopt to improve the performance of the system.

Data store drivers fetch and organize the state information in the data store leveraging the unique keys created by \sysname as described in §\ref{sec:keying_schema}.
Each data store has a specific way of organizing data: for example Cassandra organizes data in tables which can be grouped in different key spaces, while Redis has typically a single flat key space.
In our implementation, the data store driver for Redis uses the keys of \sysname directly to store and fetch state information.
For example, the key \texttt{nf1@ins1@1@Counter@abc} is used as-is to identify the counter \texttt{abc} used by core 1 of NF instance \texttt{ins1} of NF \texttt{nf1}.
Instead, the data store driver for Cassandra uses first the NF id, the NF instance id, and the core id to identify a key space.
Then, data structures of different types are stored in different tables, and the data structure id is used to identify the data structure within a table.
Using the previous example, the key space identifier is \texttt{nf1@ins1@1}, the table is \texttt{Counter}, and the id of the data structure is \texttt{abc}.
To fetch the value of the counter, we use the query \texttt{SELECT value FROM nf1@ins1@1.Counter WHERE key=abc}.

Using verbose queries and receiving bulky replies can quickly saturate the link between \sysname and the data store, ultimately decreasing the performance of the system.
For this reason, data store drivers must use the data structures offered by the data store in the most efficient way.
For example, the data store driver for Redis directly uses collections and their calls exposed by the data store.
Cassandra also supports collections, but they expose a limited number of calls, \eg it is not possible to fetch a single element from a map in an efficient manner.
For this reason, the data store driver for Cassandra implements maps by expanding them in tables, and it uses queries on tables to perform operations on maps efficiently.
As shown in Table~\ref{tab:example_queries}, every key space in Cassandra has a table ``Map'', which contains maps.
For each map, table ``Map'' contains the id of the map (in column key1) and all the key-value pairs of the map (in column key2 and column value, respectively).

\subsection{Performance Optimizations}

To decouple packet processing logic and state management, we cannot schedule the state management operations on the same threads which are processing packets.
For each Seastar thread, we create a dedicated thread to perform state management operations.
Periodically, the Seastar thread schedules the state updates for the data store to its state management thread executed in the background, while the Seastar thread keeps processing packets.
The network operator uses the configuration file to set the frequency with which Seastar threads schedule state updates.
To implement the state managements threads, we use libevent~\cite{libevent} because it integrates easily with the libraries for communicating with the data store, \ie hiredis-vip~\cite{hiredisvip} for Redis and DataStax C++ Driver~\cite{datastax} for Cassandra. 
 
\section{Evaluation}
\label{sec:evaluation}

The aim of our evaluation is to answer the following questions. 

\noindent\textbf{Does our testbed support line rate?}
\sysname is designed to process packets at line rate. 
We therefore want to make sure that the testbed we use for evaluating \sysname is able to serve packets arriving at line rate. 

\noindent\textbf{Does \sysname approach line rate?}
The goal of \sysname is to provide flexibility in changing the data store without hampering performance, thus NFs using \sysname must be able to process packets close to line rate.

\noindent\textbf{Does \sysname scale with the number of cores made available to the NF?}
NFs parallelize their packet processing across the cores made available to it, and \sysname must be able to support this and fully utilize the available resources. 

\noindent\textbf{How to quantify the benefits of performance optimizations?}
We need numerical evidence of the intuitive benefits of no\_wait calls and asynchronous updates in \sysname.

\noindent\textbf{How does the data store and its location affect \sysname?}
The decoupling between the packet processing loop and the state management should ensure that the location of the data store and the data store itself do not affect the performance of the system.
For example, we want to verify that running the data store on the same node where \sysname is running, \ie locally, or on another node, \ie remotely, does not affect the performance.

In the following, we describe our testbed (§\ref{sec:eval_equipment}) and the experiments we perform to answer to our questions (§\ref{sec:eval_results}).

\subsection{Testbed description}
\label{sec:eval_equipment}

\begin{figure}
    \centering
    \includegraphics[width=\columnwidth]{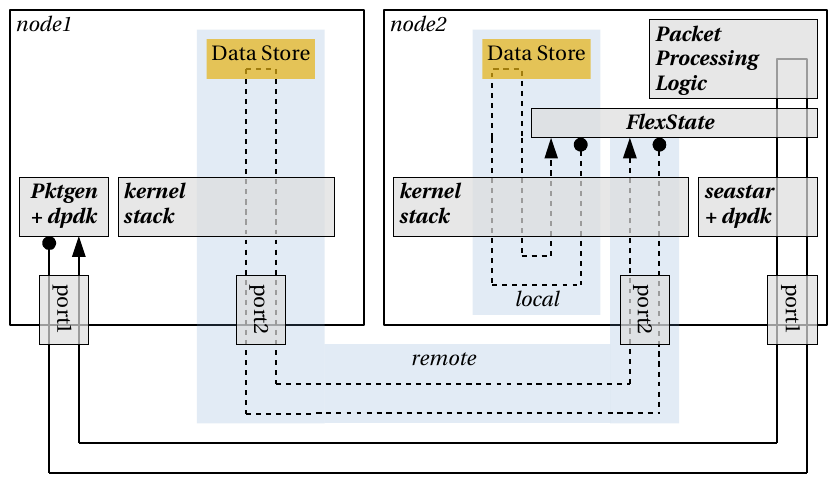}
    \caption{\textbf{Testbed workflow}.
    {\sl Pktgen sends data packets from node1 to node2 using port1, and \sysname processes the data packets before sending them back.
    Simultaneously, \sysname sends state updates to the data store, which can be local or remote.
    In the local case, the communication is confined to node2; in the remote case, \sysname communicates with the data store on node1 using port2.}}
    \label{fig:methodology}
\end{figure}

\subsubsection{Overview}
Our testbed is described in Figure~\ref{fig:methodology}.
It comprises two Dell C6320 nodes~\cite{dell}, \ie node1 and node2.
Both nodes are equipped with a Intel 82599ES 10GbE dual-port SFP+ NIC, which features two ports, port1 and port2.
The nodes are connected with a 10~Gbps link for each port pair, \ie port1 and port2 of the first node are connected to port1 and port 2 of the second node respectively.
The traffic is generated on node1 using Pktgen~\cite{pktgen}, a tool of the DPDK suite which is capable of saturating the 10~Gbps link.
The other node, node2, is used for running the NFs atop \sysname, which in turn uses Seastar and DPDK to receive and send the packets.
The first node, node1, is then used to collect the traffic processed by the NFs.
The two nodes use port1 to exchange data traffic.
If the data store is run locally, then the communication between FlexState and the data store occurs through the loopback interface; otherwise, the communication occurs through port2.

\subsubsection{Traffic Generation}
To test the support for line rate we perform our experiments in worst-case scenarios.
We generate packets of 64 bytes, which corresponds to the minimum size for a TCP packet with no payload\footnote{14 bytes of the Ethernet header, and 20 bytes each for the IP and TCP headers.}. 
The source and destination MAC addresses of the packets are set to the MAC address of the data interface of node1 and node2 respectively; the source IP address, destination IP address, source port, and destination port are generated randomly.
The generated packets are stored into a pcap file, which is then used by Pktgen as input.
The pcap file contains 50K different packets, which are sent over and over to the NFs for the whole duration of the experiment, resulting in 50K flows traversing the NFs.
In each experiment, we configured Pktgen to stream the traffic for 15 seconds.
To improve the confidence in the results, we repeat our experiments using 10 different pcap files, and the results present here are obtained computing the average of the results over all experiments.

\subsubsection{Configuring NIC and Cores}
\label{sec:nic_cores}

To test the scalability of \sysname, we run our experiments assigning to \sysname a varying number of CPU cores.
Each core is assigned a queue from the NIC port and it processes the packets arriving in that queue~\cite{Sherry15Rollback}.
In both nodes of our testbed, port1 has 16 queues~\cite{intel}, but the nodes are equipped with 48 cores.
More specifically, each node consists of two NUMA nodes, NUMA1 and NUMA2, each containing 12 physical cores, and for each physical core there is an additional virtual core due to hyperthreading.
We design and adopted a set of rules for deciding how to connect the available cores and the queues of port1.
We took into consideration the DPDK guidelines~\cite{dpdk_guidelines} that recommend to improve performance by selecting distinct cores of the same NUMA node to which the NIC is connected to, \ie NUMA2.
For this reason, we also configured the physical cores of NUMA2 with \texttt{isolcpus}, \texttt{nohz\_full}, and \texttt{rcu\_nocbs} kernel flags.
We order the 48 cores in the following manner: the twelve physical cores of NUMA2, followed by the twelve physical cores of NUMA1, the twelve virtual cores of NUMA2, and the twelve virtual cores of NUMA1; an experiment requiring $n$  cores, selects the first $n$ cores in this list. 
Note that if the number of available cores is higher than the number of queues of the NIC port, then Seastar creates software queues for the remaining available cores (§\ref{sec:impl_state}).

\subsubsection{Network Functions}
\label{sec:eval:nfs}

We consider the following NFs in our experiments.

\paragraph{testpmd}
We use testpmd~\cite{testpmd} to assess the capabilities of the testbed and to obtain a baseline for comparing \sysname's performance.
This tool of the DPDK suite performs simple operations on the packets, such as changing header information and forwarding, and it provides statistics about received, dropped, and transmitted packets.
We run testpmd on node2 by connecting it directly to port1 through DPDK, and thus skipping the software layers of Seastar and \sysname.
We configured testpmd to send back the received packets by swapping the MAC addresses.
Unlike Seastar, testpmd cannot create additional software queues, so we run testpmd using up to 16 cores only.

\paragraph{Counter NF}
To measure the impact of the software layers of Seastar and \sysname, we implement a NF which just counts the packets flowing through it.
To also measure the benefits of the asynchronous updates, we develop two versions of this NF.
In the first version, for each received packet, the NF updates immediately the counter in the data store (Sync Counter).
The second version uses asynchronous updates and \sysname updates the counter in the data store every 1~ms (Async Counter).

\paragraph{NAT and Load Balancer}
To measure the performance of \sysname with regular NFs, we implement a NAT and a load balancer using the scaffolding provided by Kablan \etal~\cite{Kablan17StatelessNF}.
The NAT substitutes source IP and source port of an incoming packet with a (IP, port) pair taken from a pool of available (IP, port) pairs.
Each flow is assigned its own pair, \ie all packets of the flow are modified using the same (IP, port) pair.
The load balancer distributes the incoming flows evenly among the servers in a given list.
When a new flow arrives to the NF, the least loaded server is selected to serve the flow. 
We adopted partitioning to implement the two NFs.
In our NAT we split the pool of available (IP, port) pairs into chunks and we assign a chunk to each core of the NF instance, while in our load balancer each core has its own load counters.
Moreover, both NFs make use of no\_wait calls and asynchronous updates, and \sysname sends state updates to the data store every 1~ms.
Note that, despite the logic of the load balancer is applied to received packets, all packets are eventually forwarded to node1.

\subsection{Experiments and Results}
\label{sec:eval_results}

\begin{figure}
    \centering
    \includegraphics[width=\columnwidth]{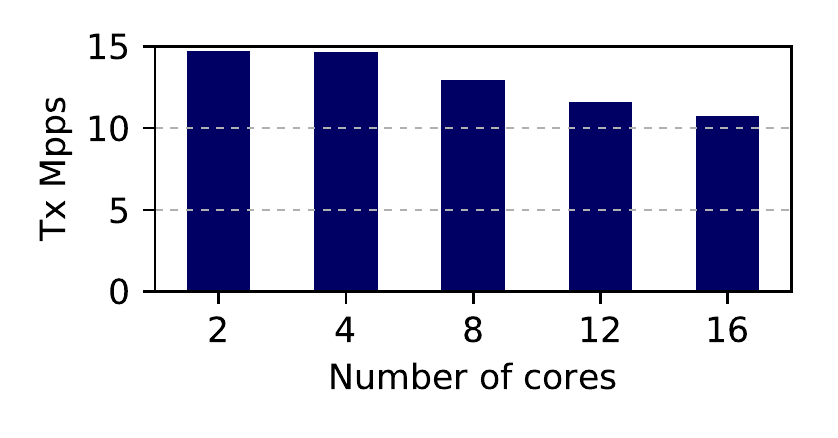}
    \caption{\textbf{Performance of testpmd}.
    {\sl The testbed supports line-rate speed, \ie 14.88 Million packets per second (Mpps).}}
    \label{fig:testpmd_only}
\end{figure}

\noindent\textbf{Does the testbed support line rate?}
We configure Pktgen on node1 to send traffic to node2 saturating the 10~Gbps link, \ie 14.88~Mpps~\cite{Duan19NetStar}.
We run testpmd on node2 and we configure it to send the received traffic back to node1.
We vary the number of cores assigned to testpmd, and this internally determines the number of NIC queues used.
Figure~\ref{fig:testpmd_only} shows the transmission rate of testpmd, \ie the number of packets forwarded back to node1 per second.
testpmd is indeed able to transmit packets back at the same rate of reception, so we can conclude that the testbed supports line-rate communication.
Note that increasing the number of assigned cores determines a deterioration in performance due to the overhead in managing additional queues~\cite{Manesh10Evaluating}.

\begin{figure*}
    \centering
    \begin{subfigure}[b]{0.48\textwidth}
        \includegraphics[width=\textwidth]{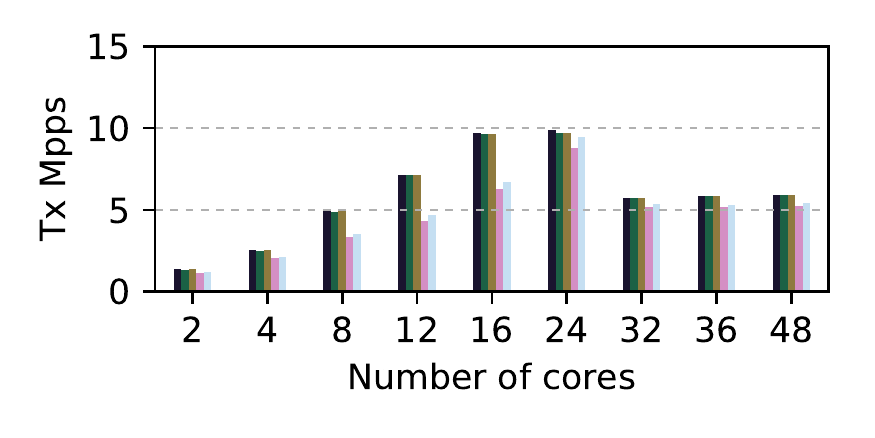}
        \caption{NAT}
        \label{fig:nat}
    \end{subfigure}
    ~
    \begin{subfigure}[b]{0.48\textwidth}
        \includegraphics[width=\textwidth]{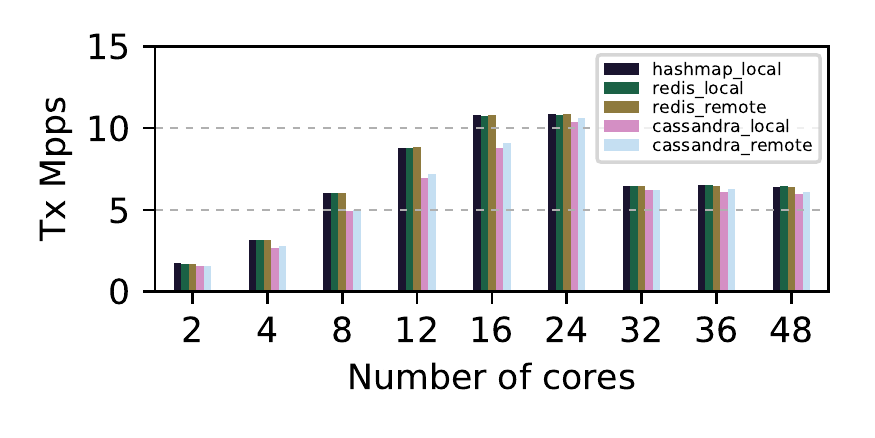}
        \caption{Load Balancer}
        \label{fig:loadbalancer}
    \end{subfigure}
    \caption{\textbf{Line-rate performance and scalability of \sysname}.
    {\sl With both NAT and load balancer, a) \sysname approaches line-rate performance, \ie close to 10~Mpps, and b) it scales with the number of cores assigned.
    Performance drops as soon as \sysname uses virtual cores, thus indicating that hypertheading is not beneficial.}}
    \label{fig:nat_loadbalancer}
\end{figure*}

\vspace{1em}
\noindent\textbf{Does \sysname approach line rate?}
Figure~\ref{fig:nat} and Figure~\ref{fig:loadbalancer} show the performance recorded in our testbed by NAT and load balancer respectively.
In particular, we measure the performance running each NF with all data stores, \ie hashmap, Redis, and Cassandra, and considering all the locations, \ie local and remote.
We can see that the NFs both record a transmission rate of about 10~Mpps when we allocate 24 cores to \sysname.
These results are in line with the performance recorded by the NFs using other state management systems~\cite{Duan19NetStar, Kablan17StatelessNF}.
We can thus conclude that \sysname is able to approach the packet processing rates of existing state management solutions. 

\vspace{1em}
\noindent\textbf{Does \sysname scale with the number of cores made available to the NF?}
The plots in Figure~\ref{fig:nat_loadbalancer} show how the NFs perform when we vary the number of cores assigned to \sysname.
In both cases, there is a steady increase in performance when going from 2 to 24 cores, which highlights the capability of \sysname to scale with the resources available.
Nevertheless, when we assign to \sysname more than 24 cores, we can see that both NFs record a drop in their performance.
We suspect this is due to the usage of virtual cores; when we assign up to 24 cores, \sysname uses distinct physical core of node2, while when we assign more than 24 cores, \sysname uses also virtual cores, as described in §\ref{sec:nic_cores}.
The physical cores have no idle time because they are busy in processing packets, thus using virtual cores forces interleaving between non-idle cores, which worsens performance.

One can note that the behaviour of testpmd is very different from the one of the two NFs, \ie testmpd performance worsen when increasing the number of cores.
We suspect that this depends on the overhead of the packet processing logic.
testpmd just performs a swap of the MAC addresses, while NAT and load balancer operate on several data structures and change several header fields before sending the packet out.
The main overhead for testpmd is therefore distributing the flows of packets to a high number of queues.
Instead, NAT and load balancer benefit from distributing the packet flows to a high number of cores because their main overhead is due to their own packet processing logic.

\begin{figure}[t]
    \centering
    \includegraphics[width=\columnwidth]{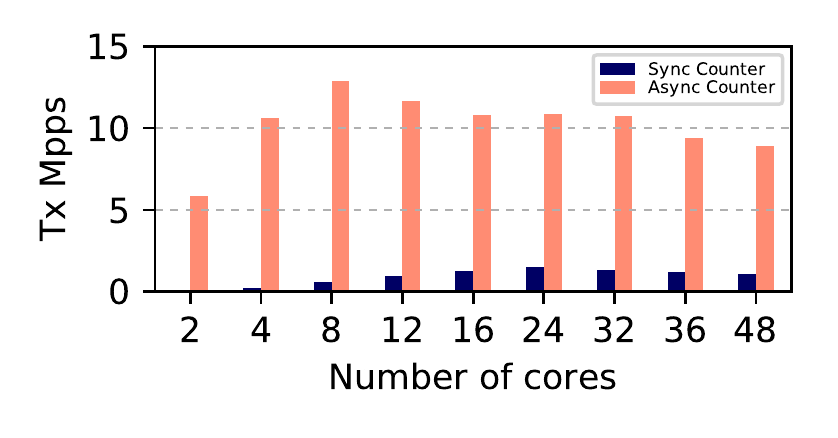}
    \caption{\textbf{Sync Counter vs Async Counter}.
    {\sl Async Counter outperforms Sync Counter in terms of transmitted packets per second.}}
    \label{fig:sync_async}
\end{figure}

\vspace{1em}
\noindent\textbf{How to quantify the benefits of performance optimizations?}
We compare the performance of Sync Counter, which does not use no\_wait calls and which communicates synchronously with the data store, with the performance of Async Counter, which uses asynchronous updates instead.
We show the comparison in Figure~\ref{fig:sync_async}.
We can see that Async Counter outperforms Sync Counter; more specifically, Async Counter reaches around 12~Mpps, while Sync Counter is never able to record more than 2~Mpps.

We can observe that Async Counter performs best when we assign 8 cores to \sysname, which confirms the need to find a trade-off between the overhead of the packet processing logic and the overhead of distributing the flows to a higher number of queues.
The packet processing logic of Async Counter only increases a counter, in addition to swapping the MAC addresses to send the packet back; the overhead of its logic is higher than the one of testpmd, but smaller than the one of NAT and load balancer.
As a result, assigning up to 8 cores benefits the performance, while the overhead of managing additional queues becomes too high when assigning more than 8 cores.

\begin{figure}[t]
    \centering
    \includegraphics[width=\columnwidth]{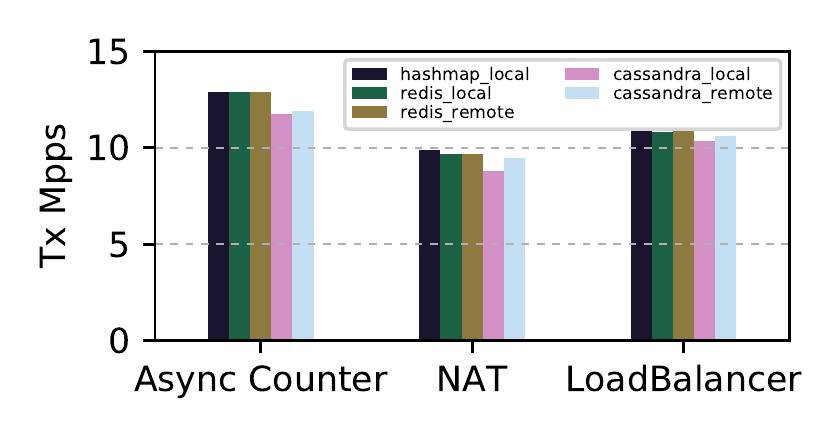}
    \caption{\textbf{Benefits of Asynchronous Updates}.
    {\sl For each NF, the values are obtained using the number of cores that achieve the best performance for the data store\_location pair.
    There are no substantial differences in performance across different combinations of data store and location.}}
    \label{fig:chart2}
\end{figure}

\vspace{1em}
\noindent\textbf{How data store and its location affect \sysname?}
Figure~\ref{fig:chart2} compares the performance of the NFs when connected with different data store and changing the location of the data store as well.
Given a data store and its location, we selected the number of cores which resulted in the NF having the best performance, and we reported the corresponding value in Figure~\ref{fig:chart2}.
We can see that a) all NFs perform close to line-rate performance, and b) for each NF, the difference in performance across different data store and different locations of the data store is negligible.
These results confirm that \sysname allows indeed the packet processing logic to operate at its own speed regardless of the data store being adopted. 
\section{Discussion}
\label{sec:discussion}

\paragraph{Consistency Tuning.}
Some state management systems support consistency tuning, \ie the NF developer can set the consistency of the state information using the API~\cite{Woo18S6}.
\sysname removes the need for this feature by applying partitioning: state information is not shared between executors, \ie cores or NF instances, so the state information is always up-to-date for each executor.

\paragraph{Locks.}
Despite the performance advantages of partitioning, some tasks may require to share state, \ie state that is accessed simultaneously by several cores or by several instances~\cite{Qazi17PEPC}.
In this case, the \sysname API can be extended to support two additional calls, \texttt{acquireLock} and \texttt{releaseLock}, which are translated by the drivers into data-store-specific mechanisms for acquiring and releasing locks respectively.
For example, both Redis and Cassandra can make use of the \texttt{IF NOT EXIST} clause to mimic a lock acquisition.
A NF could use such calls on any piece of state information to globally grant exclusive access to the state information.
Nevertheless, making use of a locking mechanism in a distributed scenario is known to be a performance killer~\cite{Khalid19CHC}.
Our recommendation is to make use of locks only as a last resort.

\paragraph{Timers.}
Some specific NFs make use of timers to carry out their tasks.
For example, when a new flow arrives, a malware detector performs a query to a registry for malware signatures, and it arms a timer to be able to react in case no reply is provided~\cite{Duan19NetStar}.
For this reason, some state management systems offer explicit support for storing timers and notifying the NF in case of expiration~\cite{Kablan17StatelessNF,Rajagopalan13SplitMerge}.
\sysname can be extended to support timers by leveraging the Time-To-Live (TTL) property offered by data stores.
\sysname associates a TTL to a record in the data store, and the NF arms the timer locally.
In this way, even in case of failure of the instance handling the timer, a new launched instance can query the data store for the timer record.
In case the record is not in the data store anymore, then the timer has expired.
Otherwise, the current TTL of the record can be used to arm a timer locally again.

\paragraph{Merging functions.}
Partitioning makes it more difficult to have a comprehensive view of the status of the network function.
For instance, when considering a load balancer, there is no single information representing the total load of the servers.
Nevertheless, this drawback can be mitigated by introducing merging functions, \ie functions that coalesce together scattered data to provide a unitary information~\cite{Woo18S6, Rajagopalan13SplitMerge, GemberJacobson14OpenNF,Peuster16Estate}.
Considering the example of the load balancer, a merging function retrieves the load of the servers for each core of each instance of the load balancer and sums them together, thus providing the user a unitary value.
Network operators can obtain a comprehensive view of the status of the network by running these functions in a cyclic fashion.

\paragraph{Multiple NF instances.}
Our evaluation shows that \sysname is able to scale with the number of cores assigned to it.
Note that \sysname is designed to scale with the number of instances as well.
Indeed, in the same way cores of a network function instance are not required to synchronize with each other, different instances of the same NF are not required to communicate with each other.
Network operators only have to assign a unique id to each instance by writing it in the configuration file.
We plan to evaluate the scalability across several instances as future work.

\paragraph{Carrier-grade data stores.}
A key advantage of \sysname is that any data store can be adopted for state management provided that the driver for it has been developed.
Instead of using ad-hoc data stores, network operators can use carrier-grade data store for managing the state of their network functions.
Carrier-grade data stores offer several advantages 
in terms of cost, maintainability and support given their typically large community of users. 
While they might not be suitable for all use-cases, we believe network operators can benefit from this possibility in specific contexts.

\paragraph{Efficiency.}
Use-cases such as Augmented Reality (AR) require both high performance and high availability~\cite{Han15NFV,Chen17ARVR}.
While we have shown that high performance can be achieved by decoupling the packet processing loop from state management, high availability can be approached increasing the frequency of the updates to the data store and mirroring the data store in multiple locations. 
At its current stage, \sysname pushes the updates to the data store without optimizations, but this can become an issue with highly frequent updates.
We are planning to complement \sysname with techniques that allow representing state changes in an efficient way~\cite{Nobach17Statelet}.

\todo{For sure reviewers are going to ask if we consider \sysname a perfect replacement of any other state management system.
The answer is no because we do not take care of aspects such as packet ordering.
This paper calls for two main contributions.
The first one is a unique interface for state management operations: in this way, network functions can be ported from one state management system to the other without need of refactoring the code.
The second one is a mechanism for adopting a different data store for state management: in this way, a network operator can change/substitute/upgrade the data store without need of changing state management system or refactoring the state management system.} 
\section{Related Work}
\label{sec:related}

\paragraph{Abstraction}
In the recent years we have witnessed how researchers have solved key problems in computer science by leveraging abstractions~\cite{Liskov09abstraction}.
Focusing on the network domain, the most glaring example is the introduction of OpenFlow~\cite{McKeown08Openflow}, which abstracts out the details of the network equipment and it provides an easy interface to network administrators.
In a similar manner, the \sysname API abstracts out the details of a single data store and it provides a unified interface with which NF developers can write the packet processing logic of NFs. The 
\sysname API resembles a Database Abstraction Layer (DBAL), a well-known and mature concept in software engineering \cite{jdo}.
Researchers and software developers have proposed several DBALs throughout the years~\cite{doctrine,prisma}, but none of them provides all the features described in Table~\ref{tab:comparison}.

\paragraph{Concurrency}
Reducing contentions between threads is critical to the performance of networked system~\cite{Katsikas18Metron}.
\sysname leverages partitioning to essentially eliminate the communication between different threads; still, there are instances in which applying a partitioning model is infeasible because of the need of having data shared between threads (§\ref{sec:discussion}).
An alternative is to apply a different concurrency model, \eg the actor model~\cite{Hewitt73Actor}, according to which the operations on a data structure are executed in strict sequence.
This guarantees that data corruption cannot occur, regardless of the thread which actually carries out the operations.
NFVActor~\cite{Duan19NFVactor} is a system for managing NFs that leverages the actor model, nevertheless the system has limited support for shared state between NF instances.

\paragraph{Other network layers}
State management is a thorny problem at every layer of the network stack.
While \sysname deals mostly with L3/L4 NFs, the problem appears in both lower, \eg L2, and upperlayers in the stack, \eg application layer, although the requirements are more homogeneous in these cases.
Systems such as EP2~\cite{Chen18EP2} and SNAP~\cite{Arashloo16SNAP} focus on relieving the NF developer from the burden of handling NF state while maintaining high performance by keeping state locally to the NFs.
Instead, availability is more relevant than performance at the application layer.
This led to the spread of data-centric paradigms, such as serverless computing, according to which applications have no own state but state is stored in a remote data store~\cite{Wang18Peeking,Hellerstein19Serverless}. 
An example of such serverless systems is Conductor from Netflix~\cite{NetflixConductor}.

\section{Conclusion}
\label{sec:conclusions}

In this paper, we have shown the significant benefits of decoupling the packet processing logic of NFs from the data store adopted for state management.
Our experiments show that decoupling can be implemented with negligible overhead in performance, and it brings two key advantages.
First, NF developers can write the packet processing logic of the NFs without being tied to the API of a specific state management system.
Second, network operators can easily change the data store used for state management, making it possible to upgrade the data store or to adopt a data store tailored for a different use-case.
 
\label{before_refs}

\bibliographystyle{ACM-Reference-Format}
\bibliography{paper}

\begin{appendix}

\section{State Identification Example}
\label{sec:keying_schema_example}
We write the packet processing logic of a simple NF which counts the total number of packets going through it.
The pseudocode of the packet processing logic using \sysname API follows:

\begin{lstlisting}[language=C++, breaklines=true, basicstyle=\footnotesize, frame=single, breakatwhitespace=true]
// create the counter
Counter pktCounter = VariableFactory::createCounter("pktCounter");

// called for every received packet
void processPacket(packet) {
    // increase the counter value
    pktCounter.add(1);
}

\end{lstlisting}

Note that the NF developer has assigned the id ``pktCounter'' to the variable.
The id also corresponds to the name the variable has in the code, but this is not required by \sysname.

A network operator who wants to run the NF compiles the following information in the configuration file fed to \sysname:

\begin{lstlisting}[breaklines=true, basicstyle=\footnotesize, frame=single, breakatwhitespace=true]
[...]
NF id: nf1;
NF instance id: ins1;
[...]
\end{lstlisting}

The network operator runs \sysname, which initiates the NF instance.
The network operator assigns a number of cores to \sysname.
Let us focus on only core 0.
\sysname needs to uniquely identify the counter being used by core 0 in the data store.
The key that \sysname builds combines the information as shown:

\begin{lstlisting}[breaklines=true, basicstyle=\footnotesize, frame=single, breakatwhitespace=true]
nf1 + ins1 + core0 + Counter + pktCounter
\end{lstlisting}

The data store driver uses the information in the key to identify the counter within the data store (§\ref{sec:impl_state}).
The schema we presented for identifying state information allows distinguishing:
\begin{itemize}
    \item state information of different NFs;
    \item state information of different NF instances;
    \item state information of different cores in the same NF instance;
    \item different data structures labeled with the same id from the NF developer.
\end{itemize}

\end{appendix} 
\label{last_page}

\end{document}